\documentclass[11pt]{article}
\usepackage{amsfonts}

\usepackage{graphics}
\usepackage{indentfirst}
\usepackage{cite}
\usepackage{latexsym}
\usepackage{amsmath}
\usepackage{amssymb}
\usepackage[dvips]{epsfig}
\usepackage{amscd}

\newtheorem{theorem}{Theorem}[section]
\newtheorem{remark}{Remark}[section]

\newtheorem{definition}{Definition}[section]
\newtheorem{lemma}[theorem]{Lemma}

\newcommand{\no}{\nonumber\\}
\newcommand{\n}{\rho}

\renewcommand{\div}{ {\rm div }  }

\newcommand{\na}{\nabla }

\newcommand{\pa}{\partial}
\renewcommand{\r}{ {R}}

\newcommand{\ia}{\int_0^T}

\newcommand{\bt}{\begin{theorem}}
\newcommand{\bl}{\begin{lemma}}
\newcommand{\el}{\end{lemma}}
\newcommand{\et}{\end{theorem}}
\newcommand{\ga}{\gamma}

\newcommand{\de}{\delta}

\newcommand{\la}{\label}

\newcommand{\p}{p(\rho)  }

\newcommand{\bn}{\begin{eqnarray}}
\newcommand{\en}{\end{eqnarray}}
\newcommand{\bnn}{\begin{eqnarray*}}
\newcommand{\enn}{\end{eqnarray*}}

\newcommand{\bnnn}{\begin{eqnarray*}}
\newcommand{\ennn}{\end{eqnarray*}}

\newcommand{\ba}{\begin{aligned}}
\newcommand{\ea}{\end{aligned}}
\newcommand{\be}{\begin{equation}}
\newcommand{\ee}{\end{equation}}
\def\O{R^3}
\def\p{\partial}
\def\norm[#1]#2{\|#2\|_{#1}}
\def\lap{\triangle}
\def\g{\gamma}
\def\ep{\varepsilon}
\def\o{\omega}
\makeatletter      
\@addtoreset{equation}{section}
\makeatother       

\title{Serrin Type Criterion for the Three-Dimensional Viscous Compressible Flows  }

\author{  Xiangdi H{\small UANG}$^{a,c},$
  Jing L{\small I}$^{b,c},$ Zhouping  X{\small IN}$^c$\thanks{This research is supported in part by Zheng Ge Ru
Foundation, Hong Kong RGC Earmarked Research Grants CUHK4040/06P and
CUHK4042/08P, and a Focus Area Grant from The Chinese University of
Hong Kong. The research of J. Li is partially supported by
 NSFC Grant No. 10971215.
 Email: xdhuang@ustc.edu.cn (X. Huang), ajingli@gmail.com (J. Li),
zpxin@ims.cuhk.edu.hk (Z. Xin).}
 \\  {\normalsize  a. Department of Mathematics,}  \\  {\normalsize  University of Science and Technology of China,} \\
  {\normalsize  Hefei 230026,
 P. R. China}
 \\
   {\normalsize  b. Institute of Applied Mathematics, AMSS,} \\   {\normalsize Academia Sinica,
  Beijing 100190,
 P. R. China  }\\
  {\normalsize   c.  The Institute of Mathematical Sciences,} \\  {\normalsize  The Chinese University of Hong Kong, Hong
  Kong}}

\date{}

\begin{document}
 \maketitle
\begin{abstract}

We extend the well-known   Serrin's    blowup criterion for the
three-dimensional (3D) incompressible Navier-Stokes equations to the
3D viscous compressible cases. It is shown that for the Cauchy
problem of the 3D compressible Navier-Stokes system in the whole
space, the strong or smooth solution exists globally if the velocity
satisfies the Serrin's condition and either the supernorm of the
density or the $L^1(0,T;L^\infty)$-norm of the divergence of the
velocity is bounded. Furthermore, in the case that either the shear
viscosity coefficient  is suitably large or there is no vacuum, the
Serrin's condition on the velocity can be removed in this criteria.

\end{abstract}


\section{Introduction}
 The time evolution of the density and the velocity of a general
viscous compressible barotropic fluid occupying a domain
$\Omega\subset R^3$ is governed by the compressible Navier-Stokes
equations
\begin{equation}\label{a1}
\left\{ \ba
& \p_t\rho + {\rm div }(\rho u)=0,\\
& \p_t(\rho u) + {\rm div }(\rho u\otimes u) -\mu\lap u-(\mu +
\lambda)\nabla({\rm div }u) + \nabla P(\rho)=0 ,\ea \right.
\end{equation}
where $\rho, u,$ and $ P$ are the density, velocity and pressure
respectively. The equation of state is given by \bnnn\la{a2}
 P(\rho) = a\rho^{\g}\quad (a>0,\g>1).
 \ennn
The constants $\mu$ and $\lambda$ are the shear viscosity and the
bulk viscosity coefficients  respectively. They satisfy the
following physical restrictions: \be \la{a9}\mu>0,\quad \lambda +
\frac{2}{3}\mu\ge 0 .\ee Let $\Omega=R^3$ and we   consider the
Cauchy problem to the equations (\ref{a1}) with initial data: \be
(\rho,u)(x,0)=(\rho_0, u_0)(x).\la{a3} \ee

 There are huge literatures on
the large time existence and behavior of solutions to (\ref{a1}).
The one-dimensional problem has been studied extensively by many
people, see \cite{Kaz,Ser1,Ser2,Hof} and the references therein. For
the    multidimensional problem (\ref{a1}),    the local existence
and uniqueness  of classical solutions are  known in \cite{Na,se1}
in the absence of vacuum and recently, for strong solutions
 also,   in \cite{K1,K3,  K2,S2} for the case where the initial density need not be positive and may
vanish in an open set. The global existence    of classical
solutions  was first  investigated by Matsumura-Nishida \cite{M1},
who proved global existence of smooth solutions for data close to a
non-vacuum equilibrium, and later by Hoff \cite{hoo,Hof2}  for
discontinuous initial data. For the existence of solutions for
arbitrary data,  the major breakthrough is due to   Lions \cite{L2}
   (see also Feireisl   \cite{F1}), where he obtains global
existence of weak solutions - defined as solutions with finite
energy - when the exponent $\gamma$ is suitably large. The main
restriction on initial data is that the initial energy is finite, so
that the density is allowed to vanish.

However, the regularity and uniqueness of such weak solutions
remains   open.  In particular,   Xin first showed    in \cite{X1}
that in the case that the initial density has compact support, any
nontrivial smooth solution to the Cauchy problem of the
non-barotropic compressible Navier-Stokes system without heat
conduction blows up in finite time for any space dimension, and the
same holds for the isentropic case (\ref{a1}), at least in
one-dimension. See also the recent generalizations to the cases for
the non-barotropic compressible Navier-Stokes system with heat
conduction (\cite{cj}) and for non-compact but rapidly decreasing at
far field initial densities (\cite{R}).

 In
this paper, we are concerned with  the main mechanism for
 possible  breakdown of strong (or smooth) solutions to the 3-D
 compressible Navier-Stokes equations.

We will use  the following conventions  throughout this paper. Set
$$\int fdx=\int_{\r^3}fdx.$$ For $1<r<\infty,$   the standard
homogeneous and inhomogeneous Sobolev spaces are denoted as follows:
   \bnnn \begin{cases}L^r=L^r(\r^3),\quad D^{k,r}  =
  \left\{u\in
L^1_{loc}(\r^3)\, \left|\norm[L^r ]{\nabla^k
u}<\infty\right\}\right.,
\quad \norm[D^{k,r} ]{u}\triangleq  \norm[L^r ]{\nabla^k u},\\
W^{k,r}  = L^r \cap D^{k,r} , \quad H^k = W^{k,2} ,\quad D^k  =
D^{k,2} ,
 \quad  D^1  = \left\{u\in L^6 \,\left|\norm[L^2 ]
 {\nabla u}<\infty \right\} \right. .\end{cases}\ennn

Next,   the   strong solutions to the Cauchy problem,
(\ref{a1})-(\ref{a3}), are defined as:
\begin{definition}[Strong solutions]  $(\rho,u)$ is called a strong solution to (\ref{a1}) in $\O\times (0,T),$ if for some $q_0\in (3,6],$\be\la{a12} \ba
& 0\le \rho\in C([0,T ],W^{1,q_0} ),\quad \rho_t\in C([0,T ],L^{q_0} ),  \\
& u\in C([0,T ],D^1\cap D^2 )\cap L^2(0,T ;D^{2,q_0} )\\
&\n^{1/2} u_t\in L^{\infty}(0,T ;L^2 ),\quad u_t\in L^2(0,T ;D^1 )
,\ea
 \ee and
$(\rho,u)$ satisfies (\ref{a1}) a.e.  in $\O\times (0,T).$
\end{definition}

 There are several recent works (\cite{H1,K1,hl,hlx1,H2,J1, H4})
 concerning   blowup criteria for   strong (or smooth)
solutions to the   compressible Navier-Stokes equations. In
particular,   it is proved in \cite{J1} for two dimensions, if
 $7\mu>9\lambda$, then\bnnn \la{a5}\lim_{T\rightarrow
 T^*}\left(\sup_{0\le t\le T}\norm[L^{\infty}]{\rho}
 +\int_0^T(\norm[W^{1,q_0}]{\rho} + \norm[L^2]{\nabla\rho}^4)dt
 \right) = \infty ,\ennn where $T^*<\infty$ is the maximal time
 of existence of a strong solution and $q_0>3$ is a constant. Later,
we \cite{H2, H4} first establish a blowup criterion, analogous to
the Beal-Kato-Majda criterion \cite{B1} for the ideal incompressible
flows, for   strong (or classical) solutions to (\ref{a1}) in three
spatial dimensions,  by assuming that if $T^*$ is the maximal time
for the existence of a strong (or classical) solution $(\n,u)$ and
$T^*<\infty,$ then  \be\la{a6} \lim_{T\rightarrow
T^*}\int_0^T\norm[L^{\infty}]{\nabla u}dt = \infty, \ee under the
  condition on viscosity coefficients: \be
\la{a7}7\mu>\lambda. \ee
 Recently, for the initial density   away from  vacuum,
  that is,   \bn\la{zq2} \inf_{x\in R^3}\rho_0(x)>0 , \en  We
\cite{hl} succeeded  in removing the crucial condition (\ref{a7}) of
\cite{H2,  H4} and established the blowup criterion (\ref{a6}) under
the physical restrictions (\ref{a9}). More recently, we \cite{hlx1}
improve the results in \cite{hl,H2,H4} by allowing vacuum states
initially and replacing (\ref{a6}) by \bn  \la{aa6}
 \lim_{T\rightarrow
T^*}\int_0^T\norm[L^{\infty}]{\mathcal{D} u}dt = \infty, \en
 where $\mathcal{D} (u)$ is
the deformation tensor: \bnn \mathcal{D}(u)  = \frac{1}{2}(\nabla u
+ \nabla u^t).\enn

Motivated by the well-known Serrin's criterion on the  Leray-Hopf
weak solutions to  the 3D incompressible Navier-Stokes equations,
which can be stated that if the velocity  $u\in L^s(0,T;L^r) $ is a
 weak solution of 3D incompressible Navier-Stokes system,
  with $r,s$ satisfying
 \bn\la{u1}
\frac{2}{s}+\frac{3}{r}\le 1 ,\quad 3<r\le\infty,\en then $u$ is
regular (see (\cite{se2,ki, bg, st}) and references therein),   we
try to extend  Serrin's blow-up criterion to
 the compressible  Navier-Stokes equations.
More precisely, we have the following  main result in this paper:

\begin{theorem}\la{t1}Let $(\rho,u)$ be a strong
solution to the Cauchy problem (\ref{a1}) (\ref{a3})
  satisfying    (\ref{a12})  while the initial data
  $(\rho_0,u_0)$ satisfy  \be\la{a10}
0\le\rho_0\in L^1  \cap H^1\cap W^{1,\tilde{q}} ,\quad u_0\in
D^1\cap D^2 , \ee for some $\tilde{q}\in (3,\infty)$ and the
compatibility condition: \be \la{a11}-\mu\triangle u_0 - (\lambda +
\mu)\nabla{\rm div }u_0 + \nabla P(\rho_0) = \rho_0^{1/2}g \quad
\mbox{for  some } g\in L^2  .\ee  If $T^*<\infty$ is the maximal
time of existence, then  both\be \la{a14}\lim_{T\rightarrow
T^*}(\norm[L^1(0,T;L^{\infty})]{\text{div}u} +
\norm[L^s(0,T;L^r)]{\rho^{\frac{1}{2}}u}) = \infty,
 \ee and \be \la{rho1}\lim_{T\rightarrow
T^*}(\norm[L^\infty(0,T;L^{\infty})]{\n} +
\norm[L^s(0,T;L^r)]{\rho^{\frac{1}{2}}u}) = \infty,
 \ee
where $r$ and $s$ satisfy   (\ref{u1}).
\end{theorem}

A few remarks are in order:
\begin{remark} If $\div u\equiv 0,$ (\ref{a14}) and (\ref{rho1}) reduce  to the
well-known Serrin's blowup criterion for 3D incompressible
Navier-Stokes equations. Therefore, Theorem \ref{t1} can be regarded
as the Serrin type blowup criterion on 3D  compressible
Navier-Stokes equations.
\end{remark}

\begin{remark} Theorem \ref{t1}  also holds for  classical
 solutions to the  3D   compressible viscous flows.
\end{remark}

\begin{remark}
  These  results can be generalized to    viscous heat-conductive
   flows, which will be reported
  in a forthcoming paper.
\end{remark}

In the following two theorems, we will show that
 (\ref{a14}) and (\ref{rho1}) can
be in fact replaced by
 \be\la{a14_2} \lim_{T\rightarrow
T^*}\norm[L^1(0,T;L^{\infty})]{\text{div}u} = \infty ,\ee and
by\be\label{rho} \lim_{T\rightarrow
T^*}\norm[L^{\infty}(0,T;L^{\infty})]{\rho} = \infty, \ee
respectively,  provided either  the viscous
  coefficients satisfy the additional condition (\ref{a7})
   besides (\ref{a9})
     or the initial
 density is away from vacuum. The first is:

\begin{theorem}\la{t3} Under the conditions of Theorem
\ref{t1}, assume that (\ref{a7})    holds in addition.  Then both
(\ref{a14_2}) and (\ref{rho})   hold  true.
\end{theorem}

\begin{remark}
  The
result in \cite{bg}  shows that   a Leray-Hopf's weak solution  to
the 3D incompressible Navier-Stokes equations becomes  smooth for
 bounded pressure. Therefore,  Corollary \ref{t3} seems reasonable since
the pressure $P$ here is bounded from above provided either
(\ref{a14_2}) or (\ref{rho}) fails. Thus Theorem \ref{t3} can be
considered as a generalization of the corresponding results in
\cite{bg}.
\end{remark}

\begin{remark}

The conclusions in Theorem \ref{t3} have been obtained independently
in \cite{wz}.
\end{remark}

 For the initial
density away form vacuum,  we have
\begin{theorem}\la{t4} In addition to the conditions of Theorem
\ref{t1}, assume that  the initial density $\rho_0$ satisfies
(\ref{zq2}). Then (\ref{a14_2}) holds.
\end{theorem}

\begin{remark} Theorem \ref{t4} improves the previous results
 in \cite{hl} and the ones in \cite{hlx1} in the absence of vacuum
  where  the criteria (\ref{a6}) and (\ref{aa6})
    have been replaced by
(\ref{a14_2}).
\end{remark}

We now comment on the analysis of this paper. The key step in
proving Theorem \ref{t1} is to derive the
$L^\infty(0,T;L^p)$-estimate on the gradient of  the density. Note
that in all previous works \cite{H2,hlx1,H4},    their methods
depend crucially on the $L^1(0,T;L^\infty)$-norm of either  the
gradient of the velocity or   its   symmetry part      instead of
the divergence. Thus, under the assumption of the left hand side of
either (\ref{a14}) or (\ref{rho1}) is finite, we need to derive the
upper bound for the $L^1(0,T;L^\infty)$-norm of the velocity
gradient. Some new ideas are needed for this. Take the case
(\ref{rho1})  for example, and assume the left hand side of   is
finite. We first obtain the estimate on the $L^\infty(0,T;L^2)$-norm
of  $\na u$ by using the a priori assumptions.
 Next we deduce the estimates on the
 $L^2(0,T;L^\infty)$-norm of both the divergence and the vorticity
 of the velocity by combining the basic  estimates on the material derivatives of the velocity
developed by Hoff \cite{hoo} and a priori estimate on
$L^\infty(0,T;L^p)$-norm of the density gradient $\na \n.$
 These estimates  can be obtained simultaneously  by
solving a logarithm Gronwall inequality based on a Beal-Kato-Majda
type inequality (see Lemma \ref{le9}) and the a priori estimates we
have just derived.

The rest of the paper is organized as follows: In Section 2, we
collect some elementary facts and inequalities which will be needed
 later.   The main results,
Theorem \ref{t1}, Corollaries\ref{t3} and   \ref{t4}  are proved in
Section 3 and Section 4 respectively.

\section{Preliminaries}\la{se2}

In this section, we  recall some  known facts and elementary
inequalities which will be used  later.

We begin with the local existence and uniqueness of strong solutions
when the initial density may not be positive and may vanish in an
open set obtained in \cite{K1}.
\begin{lemma} \la{llp}  If the initial data
$(\rho_0,u_0)$ satisfy (\ref{a10}) and (\ref{a11}), then there
exists a positive time $T_1\in (0,\infty)$ and a unique strong
solution $(\rho,u)$ to the Cauchy problem (\ref{a1})(\ref{a3})    in
$\O\times (0,T_1].$

\end{lemma}

Next, the following well-known Gagliardo-Nirenberg inequality which
will be used later frequently (see \cite{la}).

\begin{lemma}
[Gagliardo-Nirenberg]\la{l1} For  $p\in [2,6],q\in(1,\infty), $ and
$ r\in  (3,\infty),$ there exists some generic
 constant
$C>0$ which may depend  on $q,r$ such that for   $f\in H^1  $ and
$g\in L^q \cap D^{1,r} , $    we have\bn \la{g1}&&\|f\|_{L^p}^p\le C
\|f\|_{L^2}^{(6-p)/2}\|\na f\|_{L^2}^{(3p-6)/2} ,\\
\la{g2}&&\|g\|_{L^\infty} \le C
\|g\|_{L^q}^{q(r-3)/(3r+q(r-3))}\|\na g\|_{L^r}^{3r/(3r+q(r-3))} .
\en
\end{lemma}

Finally, we state the following Beal-Kato-Majda type inequality
which was proved in \cite{B1} when $\div u\equiv 0$ and will be used
later to estimate $\|\nabla u\|_{L^\infty}$ and
$\|\nabla\rho\|_{L^2\cap L^6}$.
\begin{lemma}  \la{le9}  For $3<q<\infty,$ there is a
constant  $C(q)$ such that  the following estimate holds for all
$\na u\in L^2 \cap D^{1,q} ,$ \be\la{ww7}\ba \|\na u\|_{L^\infty
}&\le C\left(\|{\rm div}u\|_{L^\infty }+ \|\na\times u\|_{L^\infty }
\right)\log(e+\|\na^2 u\|_{L^q })\\&\quad+C\|\na u\|_{L^2 } +C .
\ea\ee
\end{lemma}

{\it Proof.} The proof is similar to that of (15) in \cite{B1} and
is sketched here for completeness. It follows from  Poisson's
formula that \be\ba u (x)&=-\frac{1}{4\pi}\int \frac{\Delta
u(y)}{|x-y|}dy\\&\equiv \int  {\rm div}u(y)K(x-y)dy-\int
K(x-y)\times (\na\times u) (y)dy\\&\triangleq v+w,\ea\ee where
$$K(x-y)\triangleq\frac{ x-y }{4\pi|x-y|^3},$$ satisfies
\be \la{ww1}|  K(x-y)\le C|x-y|^{-2}, \quad |\na K(x-y)|\le
C|x-y|^{-3}.\ee

It suffices to estimate the term $\na v$ since $\na w$ can be
handled similarly (see \cite{B1}). Let $\de\in(0,1]$ be a constant
to be chosen and introduce a cut-off function $\eta_\de(x)$
satisfying $\eta_\de(x)=1$ for $|x|<\de,\eta_\de(x)=0$ for
$|x|>2\de,$ and $|\na\eta_\de(x)|\le C\de^{-1}.$ Then $\nabla v$ can
be rewritten as \be\la{ww2}\ba \na v&= \int \eta_\de(y)K(y)\na{\div
u}(x-y)dy-\int \na\eta_\de(x-y) K(x-y){\div u}(y)  dy\\& \quad+ \int
(1-\eta_\de(x-y))\na K(x-y){\div u}(y) dy  .\ea\ee Each term on the
righthand side of (\ref{ww2}) can be estimated by (\ref{ww1}) as
follows: \be\la{ww3}
\ba  & \left|\int \eta_\de(y)K(y)\na{\div u}(x-y)dy\right|\\
&\le C\| \eta_\de(y)K(y)\|_{L^{q/(q-1)}}\|\na^2u\|_{L^q}\\ &\le C
\left(\int_0^{2\de}r^{-2q/(q-1)}r^2dr\right)^{(q-1)/q}\|\na^2u\|_{L^q}\\
&\le C\de^{(q-3)/q}\|\na^2u\|_{L^q},\ea\ee \be \la{ww4}\ba
 & \left|\int \na\eta_\de(x-y) K(x-y){\div u}(y) dy\right|\\&\le
\int|\na\eta_\de(z)| |K(z)|dz \|{\div u}\|_{L^\infty} \\&\le
C\int_\de^{2\de}\de^{-1}r^{-2}r^2dr \|{\div u}\|_{L^\infty} \\&\le C
\|{\div u}\|_{L^\infty} ,\ea\ee
 \be \la{ww5}\ba  & \left|\int
(1-\eta_\de(x-y))\na K(x-y){\div u}(y) dy\right|\\&\le C\left(
\int_{\de\le |x-y|\le 1} +\int_{ |x-y|> 1} \right) |\na
K(x-y)||{\div u}(y) |dy \\&\le C \int_\de^1 r^{-3}r^2dr\|{\rm
div}u\|_{L^\infty} +C\left(\int_1^\infty r^{-6}r^2dr\right)^{1/2}
\|{\div u}  \|_{L^2}\\&\le - C  \ln \de \|{\rm div}u\|_{L^\infty} +C
\|{\na u} \|_{L^2}.\ea\ee It follows from (\ref{ww2})-(\ref{ww5})
that

\be\la{ww6} \|\na v\|_{L^\infty}\le
C\left(\de^{(q-3)/q}\|\na^2u\|_{L^q}+(1-\ln \de)\|\div
u\|_{L^\infty}+ \|\na u\|_{L^2}\right).\ee Set
$\de=\min\left\{1,\|\na^2u\|_{L^q}^{-q/(q-3)}\right\}.$ Then
(\ref{ww6}) becomes \bnn \|\na v\|_{L^\infty}\le C(q)\left(1+\ln
(e+\|\na^2u\|_{L^q})\|\div u\|_{L^\infty}+ \|\na
u\|_{L^2}\right).\enn Therefore (\ref{ww7}) holds.

\section{\la{se3}Proof of Theorem \ref{t1}}
Let $(\rho,u)$ be a strong solution to the problem
 (\ref{a1})-(\ref{a9})  as described in Theorem \ref{t1}. Then the standard
energy estimate yields \be\la{a16}\sup\limits_{0\le t\le
T}\left(\norm[L^2]{\rho^{1/2}u(t)}^2+\|\rho\|_{L^1}+\|\rho\|^\gamma_{L^\gamma}\right)
+ \int_0^T\norm[L^2]{\nabla u}^2dt \le C,\quad 0\le T<T^*. \ee

We first prove  (\ref{rho1}). Otherwise,  there exists some constant
$M_0>0$ such that \be\label{bb0} \lim_{T\rightarrow T^*}(
\|\n\|_{L^\infty(0,T;L^\infty)} + \norm[L^s(0,T;L^r)]{\sqrt{\rho}
u}) \le M_0. \ee

   The first  key estimate  on  $\nabla u $ will be
given in the following lemma.

\begin{lemma}\la{oo1}
  Under the condition (\ref{bb0}), it holds that for $0\le T<T^*,$
 \be\label{b10} \sup_{0\le t\le T}\norm[L^2]{\nabla u}^2 +
\int_0^T\int\rho u_t^2dxdt\le C, \ee where and in what follows, $C$
  denotes a generic constant depending only on $ \mu,\lambda, a,\ga,
  M_0,T,$  and the
initial data.
\end{lemma}
{\it Proof.} It follows from the momentum equations in (\ref{a1})
that \be\la{h13} \triangle G = \text{div}(\rho\dot{u}),\quad\mu
\triangle \o = \nabla\times(\rho\dot{u}), \ee where \bn \la{hj1}\dot
f\triangleq f_t+u\cdot\nabla f,\quad G\triangleq(2\mu +
\lambda)\text{div}u - P(\rho)  ,\quad\o \triangleq\na\times u, \en
are  the material derivative of $f,$ the effective viscous flux and
the vorticity respectively.

The standard $L^p$-estimate for the elliptic system (\ref{h13}),
and (\ref{g1}) give  directly that \be\label{b7} \norm[L^2]{\nabla
G} + \norm[L^2]{\nabla\o}\le C(\norm[L^2]{\rho u_t} +
\norm[L^2]{\rho u\cdot\nabla u}), \ee
  and \be\label{bb7} \norm[L^6]{\nabla G} + \norm[L^6]{\nabla\o}\le
C \norm[L^6]{\rho \dot u } \le C\|\na \dot u\|_{L^2}. \ee
 Multiplying the momentum equation $(\ref{a1})_2$ by $u_t$ and
integrating the resulting equation over $\O$ gives  \be\label{b1}\ba
& \frac{1}{2}\frac{d}{dt}\int\left(\mu|\nabla u|^2 +  (\lambda +
\mu)({\rm div}u)^2\right)dx+\int \rho u_t^2dx\\&=\int P{\rm
div}u_tdx- \int \rho u\cdot\nabla u\cdot u_tdx  . \ea\ee For the
first term on the righthand side of (\ref{b1}), one has
\be\label{b4} \ba
 & \int P{\rm div}u_tdx\\ & =
\frac{d}{dt}\int P{\rm div}udx - \int P_t{\rm div}udx\\ & =
\frac{d}{dt}\int P{\rm div}udx + \int {\rm div}(Pu){\rm div}udx +
(\g-1)\int P({\rm div}u)^2dx\\
& = \frac{d}{dt}\int P{\rm div}udx - \int (Pu)\cdot\nabla{\rm
div}udx
+ (\g-1)\int P({\rm div}u)^2dx\\
& = \frac{d}{dt}\int P{\rm div}udx - \frac{1}{2\mu+\lambda}\int
Pu\cdot\nabla Gdx -
\frac{1}{2(2\mu + \lambda)}\int P^2{\rm div}udx \\
&\quad + (\g-1)\int P({\rm div}u)^2dx\\
& \le\frac{d}{dt}\int P{\rm div}udx +\ep \|\nabla
G\|_{L^2}^2+C(\ep)\|\nabla u\|_{L^2}^2+C(\ep) ,\ea \ee due to
 \bnn  P_t + {\rm div}(Pu) + (\g-1)P{\rm div}u =
0,\enn which comes from $(\ref{a1})_1.$

For the second term on the righthand side of (\ref{b1}), Cauchy's
inequality yields

\be\label{b6} \left| \int \rho u\cdot\nabla u\cdot u_tdx \right| \le
\frac{1}{4}\int \rho u_t^2 dx+C \int \rho |u\cdot\nabla u|^2dx. \ee

Substituting (\ref{b4}) and (\ref{b6})  into (\ref{b1}), one has by
choosing $\ep$ suitably small, \bn\la{o2}
 \lefteqn{\frac{d}{dt}\int \left(\frac{\mu}{2}|\nabla u|^2 + \frac{\lambda
+ \mu}{2}({\rm div}u)^2-P{\rm div}u\right) dx+ \frac{1}{2 }\int\rho
u_t^2dx}\no&&\le C\|\nabla u\|_{L^2}^2+C \int \rho |u\cdot\nabla
u|^2dx+C.\quad\quad\quad\quad\quad\quad\en
  Gagliardo-Nirenberg's inequality  (\ref{g1}) yields that  for $r,s$
satisfying (\ref{u1}), \be\la{b8} \ba &
\norm[L^2]{\rho^{\frac{1}{2}} u\cdot\nabla u}\\ & \le
C\norm[L^r]{\rho^{\frac{1}{2}}u} \norm[L^{\frac{2r}{r-2}}]{\nabla
u}\\ \nonumber
& \le C\norm[L^r]{\rho^{\frac{1}{2}}u}(\norm[L^{\frac{2r}{r-2}}]{G} + \norm[L^{\frac{2r}{r-2}}]{\o} + 1)\\
& \le C\norm[L^r]{\rho^{\frac{1}{2}}u}(\norm[L^2]{G}^{1-\frac{3}{r}}\norm[L^2]{\nabla G}^{\frac{3}{r}} +
\norm[L^2]{\o}^{1-\frac{3}{r}}\norm[L^2]{\nabla\o}^{\frac{3}{r}} + 1)\\
&\le \ep(\norm[L^2]{\nabla G} +\norm[L^2]{\nabla\o})+ C(\ep)
\norm[L^r]{\rho^{\frac{1}{2}}u}^{\frac{s}{2}}(\norm[L^2]{G}+
\norm[L^2]{\o}+1)  + C(\ep)\\
&\le C\ep(\norm[L^2]{\rho u_t} + \norm[L^2]{\rho u\cdot\nabla u})+
C(\ep)
\norm[L^r]{\rho^{\frac{1}{2}}u}^{\frac{s}{2}}(\norm[L^2]{\nabla u}
+1)  + C(\ep) ,\ea \ee where in the last inequality we have used
(\ref{b7}).
 Thus, for $\ep$ small enough, \be\label{b9} \ba
\norm[L^2]{\rho^{\frac{1}{2}}u\cdot\nabla u}  \le C\ep
\norm[L^2]{\rho u_t} + C(\ep)
\norm[L^r]{\rho^{\frac{1}{2}}u}^{\frac{s}{2}}(\norm[L^2]{\nabla u}
+1)  + C(\ep). \ea \ee Substituting (\ref{b9})  into (\ref{o2}), one
has \bnn \lefteqn{\frac{d}{dt}\int_{\O}\left(\frac{\mu}{2}|\nabla
u|^2 + \frac{\lambda + \mu}{2}({\rm div}u)^2-P{\rm div}u\right) dx+
\frac{1}{4 }\int\rho u_t^2dx}\no&& \le C(\norm[L^r]{\rho^{1/2}
u}^s+1)(\norm[L^2]{\nabla u}^2 +1),
\quad\quad\quad\quad\quad\quad\quad\quad \enn which, together with
(\ref{bb0}) and   Gronwall's inequality, gives (\ref{b10}). The
proof of Lemma \ref{oo1} is completed.

Next, we  improve the regularity estimates on $\rho$ and $u.$
Motivated by Hoff \cite{hoo}, we start with the basic bounds on the
material derivatives of $u. $
\begin{lemma}\la{z102}
  Under the condition (\ref{bb0}), it holds that for $0\le T<T^*,$
  \be\label{le2}
\sup_{0\le t\le T}\int\rho|\dot u|^2dx + \ia\int|\nabla\dot
u|^2dxdt\le C. \ee
\end{lemma}

{\it Proof. }
 We will follow the idea  due to Hoff \cite{hoo}. Applying
  $  \dot u^j[\pa/\pa t+\div (u\cdot)]$ to
$(\ref{a2})^j$ and   integrating by parts give\be\la{m4} \ba
&  \int\rho|\dot{u}|^2dx \\
& = \int_0^t\int [ -  \dot{u}^j[\p_jP_t +\text{div}(\p_jPu)]
 + \mu \dot{u}^j[\triangle u_t^j + \text{div}(u\triangle u^j)]\\
&\quad + (\lambda+\mu) \dot{u}^j[\p_t\p_j \text{div}u  +\text{div}(u\p_j \text{div}u )]]dxds\\
& = \sum_{i=1}^{3}N_i. \ea \ee  One gets after integration by parts
and using (\ref{a1}) \be\la{m5} \ba
N_1 & = -\int_0^t\int  \dot{u}^j[\p_jP_t + \text{div}(\p_jPu)]dxds\\
& = \int_0^t\int [\p_j\dot{u}^jP^{'}\rho_t
 + \p_k\dot{u}^j\p_jPu^k]dxds\\
& = \int_0^t\int [-P^{'}\rho\text{div}u\p_j\dot{u}^j
- \p_j\dot{u}^ju^k \p_kP + \p_k\dot{u}^ju^k\p_jP ]dxds\\
& = \int_0^t\int  [-P^{'}\rho\text{div}u\p_j\dot{u}^j
+ \p_k(\p_j\dot{u}^ju^k)P - P\p_j(\p_k\dot{u}^ju^k)]dxds\\
&\le C(\int_0^t\int |\nabla u|^2dxds)^{\frac{1}{2}}
(\int_0^t\int |\nabla\dot{u}|^2dxds)^{\frac{1}{2}}\\
& \le C (\int_0^t\int
 |\nabla\dot{u}|^2dxds)^{\frac{1}{2}}. \ea \ee Integration by parts
leads to \be\la{m6} \ba
N_2 & = \int_0^t\int \mu \dot{u}^j[\triangle u_t^j + \text{div}(u\triangle u^j)]dxds\\
& = -\int_0^t\int \mu [\p_i\dot{u}^j\p_iu_t^j + \triangle u^ju\cdot\nabla\dot{u}^j]dxds\\
& = -\int_0^t\int \mu [|\nabla\dot{u}|^2 - \p_i\dot{u}^ju^k\p_k\p_iu^j - \p_i\dot{u}^j\p_iu^k\p_ku^j + \triangle u^ju\cdot\nabla\dot{u}^j]dxds\\
& = -\int_0^t\int \mu [|\nabla\dot{u}|^2 + \p_i\dot{u}^j\p_ku^k\p_iu^j - \p_i\dot{u}^j\p_iu^k\p_ku^j - \p_iu^j\p_iu^k\p_k\dot{u}^j]dxds\\
&\le -\frac{ \mu}{2}\int_0^t\int  |\nabla\dot{u}|^2dxds +
C\int_0^t\int  |\nabla u|^4dxds. \ea \ee Similarly, \be\la{m7}
 N_3  \le -\frac{\mu+\lambda}{2}\int_0^t\int
 ({\rm div} \dot u)^2dxds + C\int_0^t\int |\nabla u|^4dxds.
\ee Substituting  (\ref{m5})-(\ref{m7}) into (\ref{m4}), we obtain
immediately  by (\ref{hj1}), (\ref{bb7}), (\ref{g1}) and (\ref{g2})
that \bnn\lefteqn{ \sup_{0\le s\le t}\int\rho|\dot u|^2dx +
\int_0^t\int|\nabla\dot u|^2dxds}\no &&\le C\int_0^t\|\nabla
u\|_{L^4}^4ds+C \no &&\le C\int_0^t\left(\|G\|_{L^4}^4+\|\o
 \|_{L^4}^4\right)ds+C\no &&\le C\int_0^t\left(\|G\|_{L^2}^{5/2}
 \|\na G\|_{L^6}^{3/2}+\|\o\|_{L^2}^{5/2}
 \|\na \o\|_{L^6}^{3/2}\right)ds+C\no &&\le C\int_0^t\|\nabla
\dot u\|_{L^2}^{3/2}ds+C\no &&\le \de\int_0^t\|\nabla \dot
u\|_{L^2}^{2}ds+C_\de, \enn which  gives directly (\ref{le2}). The
proof of Lemma \ref{z102} is completed.

The next lemma is used to bound   the density gradient and
$L^1(0,T;L^\infty)$-norm of $\na u.$
\begin{lemma}\la{z3}
Under the condition $(\ref{bb0})$, it holds that for  any $
q\in(3,6] $
 \bnn\la{a48}\sup\limits_{0\le t\le
T}\left(\norm[H^1\cap W^{1,q}]{\rho}+\norm[H^1 ]{\nabla u}\right)
\le C,\quad 0\le T<T^*.\enn
 \end{lemma}

{\it Proof.} In fact, for  $  2\le p\le 6,$ $|\nabla\rho|^p$
satisfies \bnnn \ba
& (|\nabla\rho|^p)_t + \text{div}(|\nabla\rho|^pu)+ (p-1)|\nabla\rho|^p\text{div}u  \\
 &+ p|\nabla\rho|^{p-2}(\nabla\rho)^t \nabla u (\nabla\rho) +
p\rho|\nabla\rho|^{p-2}\nabla\rho\cdot\nabla\text{div}u = 0 ,\ea
\ennn which together with (\ref{g1}),  (\ref{hj1}), (\ref{bb7}) and
(\ref{le2}) gives \be\la{L11}\ba
\partial_t\norm[L^p]{\nabla\rho}&\le C(1+\norm[L^{\infty}]{\nabla u}+\norm[L^p]{\nabla G})
\norm[L^p]{\nabla\rho}\\& \le C(1+\norm[L^{\infty}]{\nabla
u}+\norm[L^2]{\nabla \dot u}) \norm[L^p]{\nabla\rho} . \ea\ee

Rewrite the momentum equations $(\ref{a1})_2$  as \bn  \la{mn0}
\mu\Delta u+(\mu+\lambda)\nabla{\rm div}u=\n \dot u+\nabla P.\en The
standard $L^p$-estimate  for the elliptic system (\ref{mn0}),
(\ref{bb7}) and (\ref{le2}) yield that for $q\in (3,6]$ \be\ba
\|\nabla u\|_{W^{1,q}}&\le C\left(\|\n\dot u\|_{L^2}+ \|\nabla
u\|_{L^2}+ \|P \|_{L^2}+\|\n\dot u\|_{L^q}+ \|\nabla
\n\|_{L^q}\right)\no   &\le C\left(1+\|\na\dot u\|_{L^2}+ \|\nabla
\n\|_{L^q}\right),\ea\ee which, combining with Lemmas \ref{l1} and
\ref{le9}, leads to\be\la{u13}
 \ba
\norm[L^{\infty}]{\nabla u} & \le C + C\left(\|\div
u\|_{L^\infty}+\|\o\|_{L^\infty}\right)
 \ln (e + \norm[W^{1,q}]{\nabla u} )  \\& \le C
 + C\left(\|\div
u\|_{L^\infty}+\|\o\|_{L^\infty}\right)
 \ln (e + \|\na\dot u\|_{L^2}  )  \\&\quad+ C\left(\|\div
u\|_{L^\infty}+\|\o\|_{L^\infty}\right) \ln (e  +\|\na \n\|_{L^q}  )
. \ea \ee Set $p=q$ in (\ref{L11}) and \bnn f(t)\triangleq e+\|\na
\n\|_{L^q},\quad g(t)\triangleq  \left(1+\|\div
u\|_{L^\infty}+\|\o\|_{L^\infty}+\|\na \dot
u\|_{L^2}\right)\log(e+\|\na \dot u\|_{L^2}) .\enn It follows  from
(\ref{L11}), (\ref{bb7}), and  (\ref{u13}) that \bnn f'(t)\le C g(t)
f(t)+ C g(t) f(t)\ln f(t)+Cg(t), \enn which yields \be\la{hb1} (\ln
f(t))'\le Cg(t)+Cg(t)\ln f(t),\ee due to $f(t)>1.$

We obtain from  (\ref{hj1}), (\ref{le2}), (\ref{g2}) and (\ref{bb7})
that \be\la{mn11}\ba &\int_0^T\left( \|\div
u\|^2_{L^\infty}+\|\o\|^2_{L^\infty}\right)dt\\ &\le C\int_0^T\left(
\|G\|^2_{L^\infty}+\|P\|^2_{L^\infty}+\|\o\|^2_{L^\infty} \right)dt
\\ &\le C\int_0^T\left(
\|G\|^2_{L^2}+\|\na G\|^2_{L^6}+\|\o\|^2_{L^2}+\|\na\o\|^2_{L^6}
\right)dt +C\\ &\le C\int_0^T  \|\na \dot u\|^2_{L^2}  dt +C\\ &\le
C,\ea\ee which together with  (\ref{le2}), (\ref{hb1})  and
Gronwall's inequality yields that \bnn \sup\limits_{0\le t\le T}
   f(t)\le C.\enn
Consequently, \bn \la{u113} \sup\limits_{0\le t\le T}\|\nabla
\rho\|_{L^q}\le C,\en
  which,  combining with  (\ref{u13}),  (\ref{mn11})  and
  (\ref{le2}),   gives directly that
\bn\la{v1} \ia \|\nabla u\|_{L^\infty}dt\le C.\en

It thus follows   from (\ref{L11}),   (\ref{le2}) and (\ref{v1})
that
  \bn \la{v2}\sup\limits_{0\le
t\le T}\|\nabla \rho\|_{L^2}\le C.\en The standard $L^2$-estimate
for the elliptic system (\ref{mn0}), (\ref{v2}) and (\ref{le2})
yield that \bn \sup\limits_{0\le t\le T}\|\nabla^2u\|_{L^2}\le C
\sup\limits_{0\le t\le T}\|\rho \dot u\|_{L^2}+C\sup\limits_{0\le
t\le T}\|\nabla\rho\|_{L^2}\le C,\en which together with
 (\ref{b10}), (\ref{a16}), (\ref{u113}), and (\ref{v2})  finishes the proof of
Lemma \ref{z3}.

The combination of Lemma \ref{z102} with  Lemma \ref{z3}   is enough
to extend the strong solutions of $(\rho,u)$ beyond $t\ge T^*$. In
fact, the functions $(\rho,u) (x,T^*)\triangleq \lim_{t\rightarrow
T^*}(\rho,u)$ satisfy the conditions imposed on the initial data
$(\ref{a10}) $ at the time $t=T^*.$ Furthermore, \bnnn -\mu\lap
u-(\mu + \lambda)\nabla({\rm div }u) + \nabla P|_{t = T^*} =
\lim_{t\rightarrow T^*} (\rho \dot u) = \rho^{\frac{1}{2}}(x,T^*) g
(x), \ennn with $g(x)\triangleq \lim_{t\rightarrow
T^*}\left(\n^{1/2}\dot u\right)(x,t)\in L^2.$  Thus,
$(\rho,u)(x,T^*)$ satisfies (\ref{a11}) also. Therefore, we can take
$(\rho,u)(x,T^*)$ as the initial data and apply Lemma \ref{llp} to
extend the local strong solution beyond $T^*$. This contradicts the
assumption on $T^*$. We thus finish the proof of (\ref{rho1}).

 It remains to prove (\ref{a14}). Assume otherwise that
\bnn \lim_{T\rightarrow T^*}(\norm[L^1(0,T;L^{\infty})]{\text{div}u}
+ \norm[L^s(0,T;L^r)]{\sqrt{\rho} u}) \le C<\infty. \enn This
together with  $(\ref{a1})_1,$  yields  immediately the following
$L^{\infty}$ bound  of the density $\rho,$ which contradicts
(\ref{rho1}). Indeed, one has
\begin{lemma}\la{o1}
  Assume that
\bnnn\la{a17}
  \int_0^T\norm[L^{\infty}]{div u}dt\le C, \quad 0< T<T^*.
\ennn Then  \be\la{a18} \sup\limits_{0\le t\le T} \norm[L^{\infty}
]{\rho} \le C,\quad 0< T<T^*. \ee Moreover, if in addition
(\ref{zq2}) holds, then \be\la{zq4} \sup\limits_{0\le t\le T}
\norm[L^{\infty} ]{\rho^{-1}} \le C,\quad 0< T<T^*. \ee
\end{lemma}
{\it Proof.} It follows from $(\ref{a1})_1$ that for $\forall p\ge
1$, \be\la{a19}\p_t(\rho^p) + {\rm div }(\rho^pu) + (p-1)\rho^p{\rm
div }u = 0. \ee Integrating $(\ref{a19})$ over $\O$ leads to
\bnnn\la{a}\p_t\int \rho^pdx  \le (p-1)\norm[L^{\infty}
]{\text{div}u}\int \rho^pdx, \ennn that is, \bnnn\la{a20}
\p_t\norm[L^p]{\rho} \le \frac{p-1}{p}\norm[L^{\infty} ]{\text{div}
u}\norm[L^p]{\rho}, \ennn
 which implies immediately
  \bnnn\la{a21}
\norm[L^p]{\rho}(t)\le C ,\ennn with $C$ independent of $p$, so
(\ref{a18}) follows. The same procedure works for $\rho^{-1}$
provided (\ref{zq2}) holds. The proof of Lemma \ref{o1} is finished.

\section{\la{se4}Proof of  Theorems \ref{t3} and   \ref{t4}}

{\it  Proof of Theorem \ref{t3}.} Theorem \ref{t3} is a consequence
of Theorem \ref{t1}, Lemma \ref{o1} and the
 following Lemma \ref{zz1}.
\begin{lemma}\la{zz1} Assume that
 (\ref{a7}) holds and (\ref{rho}) fails. Then
there exist some $q>3$ and $C>0$ such that  \be\la{z200}
  \sup_{0\le t< T^*}\int\rho|u|^q(x,t)dx \le C.
  \ee
\end{lemma}

{\it Proof.} This follows from an argument due to Hoff\cite{H3} (see
\cite{H2,H4}). Setting $q>3$ and multiplying $(\ref{a1})_2$ by
$q|u|^{q-2}u$, and integrating the resulting equation over $\O$, we
obtain by   Lemma \ref{o1} that \be\label{c1} \ba \frac{d}{dt}\int
\rho|u|^qdx + \int F dx
& = q\int {\rm div}(|u|^{q-2}u)Pdx\\
& \le C\int \rho^{\frac{1}{2}}|u|^{q-2}|\nabla u|dx\\
& \le\ep\int |u|^{q-2}|\nabla u|^2dx + C(\ep)\int \rho|u|^{q-2}dx\\
& \le\ep\int |u|^{q-2}|\nabla u|^2dx + C(\ep)(\int
\rho|u|^qdx)^{\frac{q-2}{q}}, \ea \ee for $F$ being   defined by
\be\label{c2} \ba F &\triangleq q|u|^{q-2}[\mu|\nabla u|^2 +
(\lambda + \mu)({\rm div}u)^2 + \mu(q-2)|\nabla|u||^2]
 \\&\quad+ q(\lambda + \mu){\rm div}u u\cdot  \nabla|u|^{q-2}  \\
 & \ge q|u|^{q-2}[\mu|\nabla u|^2 + (\lambda + \mu)({\rm div}u)^2 + \mu(q-2)|\nabla|u||^2\\
 & - (\lambda + \mu)(q-2)|\nabla|u||\cdot|{\rm div}u|]\\
 & = q|u|^{q-2}[\mu|\nabla u|^2 + (\lambda + \mu)({\rm div}u - \frac{1}{2}|\nabla|u||)^2]\\
 & + q|u|^{q-2}[\mu(q-2) - \frac{1}{4}(\lambda + \mu)(q-2)^2]|\nabla|u||^2\\
 & \ge C|u|^{q-2}|\nabla u|^2,
\ea \ee where we have used $|\nabla |u||\le |\nabla u|$ and the
following simple fact
   \bnn   \mu(q-1) -
\frac{1}{4}(\lambda + \mu)(q-2)^2>0, \enn due to (\ref{a7}).

Inserting (\ref{c2}) into (\ref{c1})  and taking $\ep$ small enough,
we may apply Gronwall's inequality to conclude (\ref{z200}) and thus
complete the proof of Lemma \ref{zz1}.

{\it Proof of   Theorem \ref{t4}.} Theorem \ref{t4}
   follows  from Theorem
\ref{t1} and  the next Lemma.

\begin{lemma}\label{u10} It holds that for $0<  T<T^*$, \be
\la{z201}\sup_{0\le t\le T}\int  \rho|u|^4 dx  \le C,\ee   provided
(\ref{zq2}) holds and (\ref{a14_2}) fails.
 \end{lemma}

{\it Proof.} The main idea is due to \cite{hl}. Indeed,  multiplying
$(\ref{a1})_2$ by $4|u|^{2}u$, and integrating the resulting
equation over $\O$, we obtain by using (\ref{zq4}) and (\ref{a18})
that \bn\la{o3}\lefteqn{ \frac{d}{dt}\int \rho|u|^4dx + 4\int_{\O}
 |u|^{2}\left(\mu|\nabla u|^2 +
(\lambda + \mu)({\rm div}u)^2 + 2\mu |\nabla|u||^2\right)dx}
\nonumber\\ &&=- 4(\lambda + \mu)\int
 u\cdot\nabla|u|^{2} {\rm
div}u dx + 4\int {\rm div}(|u|^{2}u)Pdx \nonumber\\
 &&\le C\int
|u|^2|\nabla u||{\rm div}u|dx+\ep \int |u|^2|\nabla u|^2dx +
 C(\ep)\int\rho|u|^2dx\nonumber\\
&&\le C\|{\rm div} u\|_{L^\infty} \left( \int \rho |u|^4dx+\|\nabla
u\|_{L^2}^2\right)+\ep \int |u|^2|\nabla u|^2dx +C(\ep).\en
Combining (\ref{o2}) with (\ref{o3}), we conclude by choosing a
positive $ \ep_0$  suitably small that \bnn
 \lefteqn{\frac{d}{dt}\int \left(\frac{\ep_0\mu}{2}|\nabla u|^2
  +\frac{\ep_0 (\lambda
+ \mu)}{2}({\rm div}u)^2-\ep_0 P{\rm div}u+\rho |u|^4\right)
dx}\no&&\quad+ \int\left(\frac{\ep_0 }{2 }\rho u_t^2 +\mu
|u|^2|\nabla u|^2\right)dx\no&&\le C\|\nabla u\|_{L^2}^2 +C\|{\rm
div} u\|_{L^\infty}\left( \int  \rho |u|^4dx+\|\nabla
u\|_{L^2}^2\right) +C, \enn which, together with Gronwall's
inequality gives (\ref{z201}). We finish the proof of Lemma
\ref{u10}.


\begin {thebibliography} {99}

\bibitem{B1} Beal, J. T., Kato, T., Majda. A.:
Remarks on the breakdown of smooth solutions for the 3-D Euler
equations.  Commun. Math. Phys. {\bf 94}, 61-66(1984)


\bibitem{bg}
Berselli, L. C., Galdi, G. P.: Regularity criteria involving the
pressure for the weak solutions to the Navier-Stokes equations.
Proc. Amer. Math. Soc. {\bf 130}(12), 3585-3595 (2002)

\bibitem{K1}  Cho, Y., Choe, H. J.,   Kim, H.:
Unique solvability of the initial boundary value problems for
compressible viscous fluid. J. Math. Pures Appl. {\bf 83},
 243-275 (2004)

\bibitem{K3} Cho, Y.,   Kim, H.:
On classical solutions of the compressible Navier-Stokes equations
with nonnegative initial densities. Manuscript Math. {\bf 120},
91-129 (2006)

\bibitem{cj} Cho, Y.,  Jin, B.J.:  Blow-up of viscous heat-conducting
compressible flows, J. Math. Anal. Appl. {\bf 320}(2),
 819-826(2006)


\bibitem{K2}  Choe, H. J.,    Kim, H.:
Strong solutions of the Navier-Stokes equations for isentropic
compressible fluids. J. Differ. Eqs. {\bf 190}, 504-523 (2003)


\bibitem{H1} Choe, H. J.,  Bum, J.:
Regularity of weak solutions of the compressible Navier-Stokes
equations.  J. Korean Math. Soc. {\bf 40}(6), 1031-1050 (2003)


\bibitem{J1} Fan, J. S.,  Jiang, S.:
Blow-Up criteria for the navier-stokes equations of compressible
fluids. J.Hyper.Diff.Equa.  {\bf 5}(1),  167-185 (2008)



\bibitem{F1} Feireisl, E., Novotny, A., Petzeltov\'{a}, H.: On the existence of globally defined weak solutions to the
Navier-Stokes equations. J. Math. Fluid Mech. {\bf 3}(4), 358-392
(2001)


\bibitem{Hof} Hoff, D.:
Global existence for 1D, compressible, isentropic Navier-Stokes
equations with large initial data. Trans. Amer. Math. Soc. {\bf
303}(1), 169-181(1987)

\bibitem{hoo}Hoff, D: Global solutions of the Navier-Stokes equations
 for
multidimensional compressible flow with discontinuous initial data.
J. Differ. Eqs.  {\bf 120}(1), 215-254(1995)

\bibitem{Hof2}Hoff, D.:
Strong convergence to global solutions for multidimensional flows of
compressible, viscous fluids with polytropic equations of state and
discontinuous initial data.  Arch. Rational Mech. Anal.  {\bf 132},
1-14(1995)


\bibitem{H3}Hoff, D.: Compressible flow in a half-space with Navier boundary
  conditions. J. Math. Fluid Mech. {\bf 7}(3), 315-338 (2005)


\bibitem{H4} Huang, X. D.:\emph{
Some results on blowup of solutions to the compressible
Navier-Stokes equations}. Ph.D Thesis. The Chinese University of
Hong Kong, 2009.


\bibitem{hl} Huang, X., Li, J.
 A Blow-up criterion for the compressible Navier-Stokes equations in
the absence of vacuum.  Methods Appl. Anal.,  in press. (2010)

\bibitem{hlx1}Huang, X. D., Li, J., Xin Z. P.:
Blowup criterion for the compressible flows with vacuum states.
Preprint




\bibitem{H2} Huang, X. D.,  Xin, Z. P.:
A blow-up criterion for classical solutions to the compressible
Navier-Stokes equations,  Sci. in China,     {\bf 53}(3),  671-686
(2010)

\bibitem{Kaz} Kazhikhov, A. V.,  Shelukhin, V. V.:
Unique global solution with respect to time of initial-boundary
value problems for one-dimensional equations of a viscous gas.
Prikl. Mat. Meh.  {\bf 41}, 282-291 (1977)

\bibitem{ki}
Kim, H.: A blow-up criterion for the nonhomogeneous incompressible
Navier-Stokes equations.  Siam J. Math. Anal. {\bf 37}(5), 1417-1434
(2006)

\bibitem{la}
Ladyzenskaja, O. A., Solonnikov,  V. A.,   Ural'ceva, N. N.: Linear
and quasilinear equations of parabolic type, American Mathematical
Society, Providence, RI (1968)


\bibitem{L2}  Lions, P. L.:
\emph{Mathematical topics in fluid mechanics}.
 Vol. {\bf 2}. Compressible models. New York: Oxford
University Press, 1998


\bibitem{M1} Matsumura, A.,   Nishida, T.:  The initial value problem for the equations of motion of viscous and heat-conductive
gases. J. Math. Kyoto Univ. {\bf 20}(1), 67-104 (1980)


\bibitem{Na} Nash, J.: Le probl\`{e}me de Cauchy pour les \'{e}quations
diff\'{e}rentielles d'un fluide g\'{e}n\'{e}ral. Bull. Soc. Math.
France. {\bf 90},487-497 (1962)


\bibitem{R}Rozanova, O.:  Blow up of smooth solutions to the compressible
Navier-Stokes equations with the data highly decreasing at infinity,
J. Differ. Eqs.  {\bf 245},  1762-1774 (2008)

\bibitem{S2} Salvi,R.,  Straskraba, I.:
Global existence for viscous compressible fluids and their behavior
as $t\rightarrow \infty$. J. Fac. Sci. Univ. Tokyo Sect. IA. Math.
{\bf 40}, 17-51 (1993)

\bibitem{Ser1} Serre, D.:
Solutions faibles globales des \'equations de Navier-Stokes pour un
fluide compressible. C. R. Acad. Sci. Paris S\'er. I Math.
 {\bf 303}, 639-642 (1986)

\bibitem{Ser2} Serre, D.:
Sur l'\'equation monodimensionnelle d'un fluide visqueux,
compressible et conducteur de chaleur. C. R. Acad. Sci. Paris S\'er.
I Math. {\bf 303}, 703-706 (1986)

\bibitem{se1} Serrin, J.: On the uniqueness of compressible fluid motion,
Arch. Rational. Mech. Anal. {\bf 3}, 271-288 (1959)

\bibitem{se2} Serrin, J.: On the interior regularity of weak solutions of the {N}avier-{S}tokes equations
Arch. Rational Mech. Anal.,{\bf 9}, 187-195 (1962)

\bibitem{st} Struwe, M.: On partial regularity results for the Navier-Stokes equations, Comm. Pure
Appl. Math. {\bf 41},  437-458(1988)

\bibitem{wz} Sun, Y. Z., Wang, C., Zhang, Z. F.
A Beale-Kato-Majda Blow-up criterion for the 3-D compressible
Navier-Stokes equations, (2010) Preprint.

\bibitem{X1}Xin, Z. P.:
Blowup of smooth solutions to the compressible {N}avier-{S}tokes
equation with compact density. Comm. Pure Appl. Math.   {\bf 51},
229-240 (1998)


\end {thebibliography}
\end{document}